\documentclass[12pt]{article}
\usepackage{latexsym}
\usepackage{amsfonts}
\usepackage{epsfig}
\usepackage{amssymb}
\usepackage{graphicx}

\begin{document}
\bigskip\bigskip
\begin{center}
{\bf{\large THE VISCOSITY BOUND CONJECTURE}}
\end{center}
\begin{center}
{\bf{\large AND}}
\end{center}
\begin{center}
{\bf{\large HYDRODYNAMICS OF M2-BRANE THEORY}}
\end{center}
\begin{center}
{\bf{\large AT FINITE CHEMICAL POTENTIAL}}
\end{center}
\vfill
\renewcommand{\thefootnote}{\fnsymbol{footnote}}
\centerline{{\large Omid
Saremi}\footnote{omidsar@physics.utoronto.ca}}
\bigskip\centerline{\it{ McLennan Physical Laboratories }}
\centerline{\it{University of Toronto, 60 St George Street, }}
\centerline{\it{Toronto, ON, Canada, M5S 1A7}}
\vfill
\begin{abstract}
Kovtun, Son and Starinets have conjectured that the viscosity to
entropy density ratio $\eta/s$ is always bounded from below by a
universal multiple of $\hbar$ i.e., $\hbar/(4\pi k_{B})$ for all
forms of matter. Mysteriously, the proposed viscosity bound appears
to be saturated in all computations done whenever a supergravity
dual is available. We consider the near horizon limit of a stack of
M2-branes in the grand canonical ensemble at finite R-charge
densities, corresponding to non-zero angular momentum in the bulk.
The corresponding four-dimensional R-charged black hole in Anti-de
Sitter space provides a holographic dual in which various transport
coefficients can be calculated. We find that the shear viscosity
increases as soon as a background R-charge density is turned on. We
numerically compute the few first corrections to the shear viscosity
to entropy density ratio $\eta/s$ and surprisingly discover that up
to fourth order all corrections originating from a non-zero chemical
potential vanish, leaving the bound saturated. This is a sharp
signal in favor of the saturation of the viscosity bound for event
horizons even in the presence of some finite background field
strength. We discuss implications of this observation for the
conjectured bound.
\end{abstract}
\vfill

\noindent January 20th, 2006.

\setcounter{page}{0}

\newpage
\section{\small Introduction}

In recent years, string theory/gauge theory dualities have been
proven to be immensely invaluable tools in understanding strong
dynamics of certain quantum field theories. For any strong-weak type
duality, computations that are possible on both sides of the duality
are scarce. That is where taking various limits of both sides of the
duality comes in handy, and may provide evidence for this in the
past include the PP-wave limit of the AdS/CFT correspondence
\cite{PPwave} among others. The {\em hydrodynamic limit} could be
one such simplifying yet non-trivial limit where even highly quantum
mechanical systems behave simply and universally. Hydrodynamics
appears to be relevant to achieve a better understanding of the
quark-gluon plasma (QGP) state in heavy ion collisions at RHIC.
Studying near equilibrium phenomena in a hot, strongly coupled QCD
plasma is never easy, even on the lattice. Extracting retarded
Green's functions from Euclidean lattice computations requires long
Minkowski time separations, which calls for a large number of
lattice points \cite{Lattice}. Therefore using a dual description
for QCD like theories in order to extract transport coefficients may
find even practical use in studying near-equilibrium QCD physics.

Sometime ago, Policastro, Son and Starinets proposed a
prescription for calculating Minkowskian field theory Green's
functions using the supergravity dual \cite{Recipe}. Since then an
extensive study has been done where, using this prescription,
various transport coefficients were calculated from the gravity
side corresponding to the D-brane world-volume theories
\cite{WholeBunch} as well as M2 and M5-brane theories in M-theory
\cite{Herzog}. In \cite{Conjecture}, it was conjectured that the
ratio of shear viscosity to entropy density is bounded from
below\begin{equation}\frac{\eta}{s}\geq\frac{\hbar}{4\pi
k_{B}}=6.08\times10^{-13} K.s,
\end{equation} where $\eta$ is the shear
viscosity and $s$ is the entropy density. The bound is well
satisfied for weakly coupled systems, which could be understood
intuitively by noticing that in a weakly coupled plasma, mean free
flight time for the constituents is long. The puzzling feature
shared by every transport coefficient calculation performed so far
\cite{AlexLiuUniversality} is that for all known holographic duals
to various supersymmetric gauge theories at finite temperature, the
proposed bound appears to be {\em saturated} \cite{WholeBunch}. This
suggests that, at the
 {\em infinite coupling} limit where the supergravity description is adequate, there exists
some sort of {\em universality} in the hydrodynamic description of
all of these field theories (their dual being some black hole in
anti-de Sitter space) \cite{Absorption}. It was also emphasized
\cite{Absorption} that the universal nature of the ratio is
connected to the universality of the black hole absorption cross
section for low energy graviton scattering, at least for the cases
where the AdS near horizon region has a flat space completion.

Just on dimensional grounds, the bound itself appears to be in
harmony with the observation that $\eta/s$ is a product of the
energy per effective degree of freedom in the field theory at a
t'Hooft coupling $g_{YM}^2N\gg 1$ and a time scale associated with
the mean free flight time of the quasi-particle excitation
\cite{Absorption}. According to the Heisenberg uncertainty
principle, this product must be bounded from below by a multiple
of $\hbar$ in order for the notion of quasi-particle to make
sense. In \cite{AlexUniversality}, the universal nature of the
ratio was further established and extended to a large class of
supergravity backgrounds where the dual possesses a
translationally invariant horizon. The entire class {\em
saturates} the bound.

There is no known example where, at the infinite coupling limit, the
bound holds but {\em isn't saturated}. One major motivation behind
this work was to investigate whether there exist example(s) in
which, at the strict infinite coupling limit, the viscosity bound is
satisfied but not saturated. A particular setup which could
potentially avoid the no-go theorems discussed in
\cite{AlexUniversality}, involves supersymmetric gauge theories
living on the world-volume of type II D-branes or membrane and
five-brane theories in M-theory at finite global charge densities.
From a lower dimensional prospective this corresponds to black holes
with some finite gauge field strength turned on at the horizon. It
is natural to believe that the hydrodynamic properties of these
horizons with a non-zero gauge field must be different from the
"{\em neutral horizons}".

Here we work in the grand canonical description at finite chemical
potential corresponding to finite R-charge i.e., $\langle
j^{0}\rangle\neq 0$ where $j^{0}$ is the R-charge. Gravitationally,
the R-charge arises from finite transverse rotation from a 10 or
11-dimensional point of view.

We find that turning on a finite  R-charge background {\em
increases} the viscosity. In the conclusion section we speculate as
to why this happens. To our surprise, we discover that the viscosity
to entropy density ratio $\eta/s$, remains the same as that for zero
chemical potential case i.e., $1/(4\pi)$ up to forth order in powers
of $\Omega/T_{H}$, where $\Omega$ is the angular velocity and
$T_{H}$ is the Hawking temperature. This provides clear evidence for
the saturation of the bound even for the horizons with a finite
gauge field.

The paper is organized as follows. We begin with a lightning review
of the hydrodynamic limit in systems with many degrees of freedom.
After reviewing the prescription for computing Minkowskian retarded
Green's functions in chapter 3. The relevant gravitational
background, i.e., R-charged Anti- de Sitter black hole is discussed
in chapter 4. Gravitational perturbation theory of AdS$_{4}$ black
holes is the subject of discussioin in chapter 5. In chapter 6, We
consider $\langle T_{\mu\nu}T_{\rho\sigma}\rangle$ correlation in
Minkowskian signature using the prescription reviewed in chapter 3.
Finally, we proceed to calculate the transport coefficient of
interest here namely the shear viscosity $\eta$ and $\eta/s$ in
chapter 7.

\section{\small Hydrodynamic Limit of Quantum Field Theories}

Linear response theory is the mathematical theory of the relaxation
of small disturbances around equilibrium where the thermally
averaged Minkowskian Green's functions (retarded, in order to
account for causality) of the unperturbed system fully characterize
the system's response to the external stimuli
\begin{eqnarray}\label{RetardedGF}
G_{\mu_{1}\ldots\mu_{j}\ldots}^{R}(\vec{q})&=&\int d^{d}x e^{-iq.
x}\theta(t)\langle[
\widehat{Q}_{\mu_{1}\ldots}(x),\widehat{Q}_{\mu_{j}\ldots}(0)
]\rangle_{\beta}.
\end{eqnarray}
Here, $G_{\mu_{1}\ldots\mu_{j}\ldots}^{R}(q)$ denotes the retarded
Green's function, $Q_{\mu_{1}\ldots}(x)$ is the operator
corresponding to the conserved current which couples to the
external world disturbance,
 $\mu_{i}$s are some spacetime indices and $\beta$ is the inverse
temperature. It is argued, and in simple cases explicitly shown,
that the slowly varying (both in space and time) behavior of the
Minkowskian Green's functions of interacting field theories has a
specific pole structure imposed by the ``hydrodynamic equations''.
These hydrodynamic limit conditions are usually satisfied when local
thermal equilibrium is achieved. To create such circumstances, a
fluid must be in its high collision regime where interactions are
important to the dynamics. Hydrodynamics is the study of small, long
wavelength and low frequency fluctuations of a medium in the
vicinity of its equilibrium point\footnote{Clearly, in a conformal
theory at finite temperature, the temperature provides the required
length scale for the system.} In this limit all the fine structure
of cutoff-scale physics gets wiped out, leaving only a few transport
coefficients at low energies and long distances. It turns out that
the relevant degrees of freedom to a hydrodynamical description are
the charge densities of various global symmetries at the UV cutoff
scale, along with the phase of the order parameters if any phase
transition exists \cite{Kovtun}. This is in accord with the fact
that relaxation of any disturbance in conserved densities, in the
deep IR limit, diverges. This makes charge densities the only
relevant degrees of freedom in the hydrodynamic limit. Transport
coefficients appearing in these sets of equations are not themselves
part of the hydrodynamic description but, rather, are inputs. These
coefficients could, in principle, be calculated from the slow
varying part of the two point Green's functions like
(\ref{RetardedGF}). It is exactly this type of computations which we
will be concerned with in this paper.

 A fluid in equilibrium has a spatial energy-momentum tensor
of the form $T_{ij}=p\delta_{ij}$. Slightly away from the
equilibrium, extra stresses will be present as a result of viscous
forces. The viscous part of the energy-momentum tensor of a fluid
is proportional to the symmetric-traceless combinations of the
momentum gradient of the fluid. It takes the following form
\begin{eqnarray}\label{EnergyMomentumTensor}
T_{ij}&=&\delta_{ij}p-\eta(\frac{\partial{\pi_{i}}}{\partial{x_{j}}}
+\frac{\partial{\pi_{j}}}{\partial{x_{i}}})-(\zeta-\frac{2}{3}\eta)\delta_{ij}\nabla{\vec{\pi}},
\end{eqnarray}
where $\eta$ denotes the shear viscosity and microscopically is
inversely proportional to the mean scattering rate. It is directly
proportional to the mean free path for the effective degrees of
freedom at a given value of the coupling. A strongly coupled fluid
will have less viscosity as the mean free path is small and the
energy in the perturbation taking the system away from its
equilibrium, gets redistributed among degrees of freedom very
quickly. This simply means that strongly coupled fluids are better
approximations to the hypothetical notion of an ``{\em ideal
fluid}".

Also, $\zeta$ is the bulk viscosity, $p$ is the local momentum
density, and $\vec{\pi}$ is the momentum flux. In a conformal
theory, $\zeta=0$. The energy-momentum tensor
(\ref{EnergyMomentumTensor}), along with the conservation equation
for the energy-momentum tensor
\begin{eqnarray}\label{Conservation}
\partial_{t}T^{0i}+\partial_{j}(T^{ij}-p\delta^{ij})&=&0,
\end{eqnarray}
\begin{eqnarray}
\label{Conservation}
\partial_{t}T^{00}+\partial_{i}T^{0i}&=&0,
\end{eqnarray}
form the complete set of the hydrodynamic equations specifying the
system in its hydrodynamic limit. The natural frequency of the
system specified by (\ref{EnergyMomentumTensor}) and
(\ref{Conservation}), which we will be focusing on in this paper,
is called the shear mode and is shown to possess the following
dispersion relation
\begin{eqnarray}\label{Dispersion}
\omega&=& -i\mathcal{D}q^2,
\end{eqnarray}
where $\mathcal{D}=\eta/(\epsilon+p)$ is called diffusion constant,
$\epsilon$ is the internal energy density of the system and $p$ is
the pressure. Noticing that the dispersion relation
(\ref{Dispersion}) appears as the pole structure of $G^{R}(\vec{q})$
in the hydrodynamic limit, enables us to read off the transport
coefficient $\eta$.
\section{\small AdS/CFT in Minkowski Signature}

In order to explore the hydrodynamic limit, one has to find a way to
calculate the two point functions in Minkowskian signature. The
celebrated Anti-de Sitter/Conformal Field Theory duality
\cite{Maldacena} is naturally formulated in Euclidean signature
where the boundary configuration of the on-shell closed string
background in the bulk of AdS acts as a source for generating all
the Green's functions of the corresponding boundary operator.
Formally, one can write
\begin{eqnarray}
\langle\mathcal{O}(x_{i_{1}})\mathcal{O}(x_{i_{2}})\ldots\mathcal{O}(x_{i_{n}})\rangle\propto\frac{\delta
^{n}S_{cl}[\Phi_{cl}(x)]}{\delta\Phi_{B}(x_{i_{1}})\ldots\delta\Phi_{B}(x_{i_{2}})\delta\Phi_{B}(x_{i_{n}})},\\\nonumber
\end{eqnarray}
where $\Phi_{B}(x_{i})$ represents the closed string boundary
value associated to the operator $\mathcal{O}(x)$ in the boundary
theory and $S_{cl}$ is the on-shell action. Calculating
Minkowskian Green's functions is blocked due to the
characteristics of the boundary value problem for hyperbolic
operators in Minkowskian signature AdS spacetime. The authors in
\cite{Recipe} argued for a Minkowskian prescription for computing
the Minkowskian thermal correlators in the boundary theory.
According to this prescription
\begin{eqnarray}\label{Prescription}
G_{R}(\vec{q})&=&-2\mathcal{F}(u,\vec{q})|_{u=u_{boundary}},
\end{eqnarray}
where $u$ is the radial coordinate in AdS to be defined later,
$\vec{q}$ is the momentum on the boundary, and the imaginary part
of $\mathcal{F}$ is the Fourier component of the flux associated
with the corresponding AdS fluctuation. The contribution at the
horizon is discarded from (\ref{Prescription}). Moreover only
incoming solutions to the AdS equations of motion are kept at the
horizon. The prescription was further examined and confirmed to
reproduce the desired and known results for the two examples
worked out in \cite{Recipe} i.e., zero temperature
$\mathcal{N}=4$,  SYM theory in four dimensions and 2-dimensional
finite temperature CFT dual to the BTZ black holes in $AdS_{3}$.
To see the detail of the definitions and justifications refer to
\cite{Recipe}.
\section{\small Gravitational Background}

\subsection{\small $D=4$, $\mathcal{N}=2$ U(1)$^4$ Extended Gauged Supergravities}

Extended gauged supergravities arise as Kaluza-Klein reductions of
both D=10 and D=11 supergravities. Amongst these
compactifications, an $S^7$ reduction of D=11 supergravity to N=8
SO(8) gauged supergravity in D=4 admits a consistent truncation
where only gauge fields in the Cartan subalgebra of the gauge
group SO(8), i.e., a $U(1)^4$ subgroup of the gauge group,
survive. This truncation allows for 4-charge AdS black holes in
four dimensions. Such a consistent truncation of D=4, N=8 SO(8)
supergravity with $U(1)^4$ gauge group includes four commuting
gauge fields, three dilatons, and three axions and is called
minimal N=2, $U(1)^4$ supergravity. The Lagrangian for N=2 minimal
supergravity is given by
\begin{eqnarray}
e^{-1}\mathcal{L}&=&R-\frac{1}{2}(\partial\vec{\phi})^2+8g^2(\cosh{\phi_{1}}+\cosh{\phi_{2}}+\cosh{\phi_{3}})\\\nonumber
&&-\frac{1}{4}\sum_{i=1}^{4}e^{\vec{a_{i}}.\vec{\phi}}(F^{i}_{(2)})^2.
\end{eqnarray}
where the $\phi_{i}$'s are 3 dilatons, $g$ is the inverse
$AdS_{4}$ radius. The $\vec{a}_{i}$ are introduced in
\cite{Cvetic}.
 It can be shown that any non-axionic solutions to this minimal supergravity can be uplifted to
11-dimensional supergravity using the ansatz presented in
\cite{Cvetic}.

\subsection{\small 4-Charge AdS$_{4}$ Black Holes}

The near horizon geometry of the non-extremal M2-brane background
which allows for up to four independent angular momenta (the
dimension of the Cartan subalgebra of SO(8)) is given by
\cite{Cvetic}
\begin{eqnarray}\label{RotatingM2Brane}
ds_{11}^2&=&\tilde{\Delta}^{2/3}\large[-(H_{1}H_{2}H_{3}H_{4})^{-1/2}fdt^2+(H_{1}H_{2}H_{3}H_{4})^{1/2}(f^{-1}dr^2+r^2d\vec{y}.d\vec{y})\large]\\\nonumber
&&+g^{-2}\tilde{\Delta}^{-1/3}\sum_{i=1}^{4}X_{i}^{-1}(d\mu_{i}^2+\mu_{i}^2(d\phi_{i}^2+gA^{i}_{t}dt)^2),
\end{eqnarray}
where
\begin{eqnarray}
f&=&-\frac{\mu}{r}+4g^2r^2H_{1}H_{2}H_{3}H_{4},\\\nonumber
H_{i}&=&1+\frac{l_{i}^2}{\tilde{r}^2}, i=1,2,3,4,\\\nonumber
X_{i}&=&H_{i}^{-1}(H_{1}H_{2}H_{3}H_{4})^{1/4},\\\nonumber
A_{t}^{i}&=&\frac{1-H_{i}^{-1}}{g l_{i}\sinh{\alpha}},\\\nonumber
\tilde{\Delta}&=&\sum_{i=1}^{4}X_{i}\mu_{i}^2,\\\nonumber
r&=&\frac{1}{2}(2m\sinh^2{\alpha})^{-1/6}\tilde{r}^2,\\\nonumber
g^2&=&(2m\sinh^2{\alpha})^{-1/3},\nonumber
\end{eqnarray}
where the $\mu_{i}$'s are coordinates on the unit 3-sphere and
$l_{i}$ are four angular momentum parameters. The decoupled
background (\ref{RotatingM2Brane}) gives rise to a duality between
$AdS_{4}\times S^{7}$ and the M2-brane theory at finite temperature
and finite R-charge, such that in thermal equilibrium $\langle
j^{0}\rangle\neq 0$, where $j^{\mu}$ represents the R-charge current
in the boundary theory.

The near horizon limit of the rotating M2-branes background
(\ref{RotatingM2Brane}) under an $S^7$ reduction gives rise to the
four dimensional, 4-charge AdS black holes. As a result, it is
more convenient to work within the framework of N=2 minimal
supergravity described above while bearing in mind that any
solution to this theory is an M-theory solution. Therefore we will
focus only on constructing minimal supergravity perturbations.

To simplify the setup without losing much of the generality, we
choose to work with black holes with four equal charges which
further simplifies the N=2 minimal supergravity to just Einstein
gravity with a cosmological constant coupled to four Maxwell fields
in four dimensions. In fact, setting four charges equal makes the
dilaton sector decouple and partially simplifies the system without
destroying its essential features.

AdS$_{4}$ black holes with four equal charges in this theory, with
the horizon geometry being a space of constant curvature, are given
by \cite{Cvetic}
\begin{eqnarray}\label{AdS4BH}
ds_{4}^2&=&-H^{-2}fdt^2+H^{2}(f^{-1}dr^2+r^2d\Omega_{2,k}^2),\\\nonumber
f&=&k-\frac{\mu}{r}+4g^2r^2H^4,\\\nonumber
H&=&1+\frac{\mu\sinh^2{\beta}}{kr},\\\nonumber
A^{i}_{t}&=&\sqrt{k}(1-H^{-1})\coth{\beta}, i=1,2,3,4,\nonumber
\end{eqnarray}
where $k=-1, 0, 1$ refers to the curvature of the horizon
geometry.\footnote{ Note that in spacetimes with AdS$_{4}$
asymptotics, horizon topology is not restricted to just
2-spheres:the horizon manifold could be either of the 3 possible
spaces of constant curvature.}

The case $k=0$ needs special treatment and the result is the same
background except the gauge field and the function $H$ are now
changed to
\begin{eqnarray}
H&=&1+\frac{\mu\sinh^2{\beta}}{r},\\\nonumber
A_{t}^{i}&=&\frac{1-H^{-1}}{\sinh{\beta}}, i=1,2,3,4.\nonumber
\end{eqnarray}
The relation between the eleven and four dimensional metric is
\begin{eqnarray}
l_{i}^2g&=&2\mu\sinh^2{\beta_{i}},\\\nonumber
\sinh\beta_{i}&=&gl_{i}\sinh{\alpha},\\\nonumber
\mu&=&mg^5.\nonumber
\end{eqnarray}

In what follows, we choose to work with the flat case $k=0$. Upon
uplifting to M-theory, this gives rise to the decoupling limit of
the flat world-volume (parameterized by coordinates on
$\Omega_{2,0}$) rotating M2-branes with all four possible rotation
parameters going. Note that the metric now is written as
\begin{eqnarray}
ds_{4}^2&=&-H^{-2}fdt^2+H^2(f^{-1}dr^2+r^2(d\theta^2+d\phi^2)),
\end{eqnarray}
where $\theta$ and $\phi$ are the dimensionless angular variables.
\footnote{The prescription for switching to the dimensionful
coordinates $x$,$y$ used in \cite{Herzog} is to notice that
$2gx=\theta$ and $2gy=\phi$.}

Let us introduce a more commonly used \cite{Herzog} ``$u$''
coordinate for later use
\begin{eqnarray}\label{defHf}
r&=&\frac{R_{0}^2}{u},\\\nonumber
R_{0}^6&=&\frac{\mu}{4g^2},\nonumber
\end{eqnarray}
where $u$ is the new membrane radial coordinate. Note that functions
$f$ and $H$ appearing in (\ref{AdS4BH}) are now written as
\begin{eqnarray}\label{MasterFEq}
f&=&\frac{4g^2R_{0}^4}{u^2}(H^4-u^3),\\\nonumber
H&=&1+\frac{\mu\sinh^2{\beta}u}{R_{0}^2}.\nonumber
\end{eqnarray}
Here, we record a few quantities associated to the background
(\ref{RotatingM2Brane}) for later usage.

A dimensionless combination $y=\mu \sinh^2{\beta}/R_{0}^2$ will make
an appearance later on. The angular velocity $\Omega$ corresponding
to (\ref{RotatingM2Brane}) is proportional to the chemical potential
for the R-charge as viewed from the dual boundary theory. To
calculate $\Omega$, we need to rewrite the M-theory embedding of our
$AdS_{4}$ black holes with four equal charges namely
(\ref{RotatingM2Brane})\cite{Cvetic}
\begin{eqnarray}
ds_{11}^2&=&-H^{-2}fdt^2+H^2(f^{-1}dr^2+r^2d\vec{y}.d\vec{y})+\sum_{i=1}^{4}(d\mu_{i}^2+\mu_{i}^2(d\phi_{i}+gA_{t}^{i}dt)^2).
\end{eqnarray}
Using definition of $\Omega$
\begin{eqnarray}
\Omega_{i}&=&-\frac{g_{t\phi_{i}}}{g_{\phi_{i}\phi_{i}}},
\end{eqnarray}
where ``$i$'' labels each of the four independent angular
velocities. One obtains
\begin{eqnarray}
\Omega_{i}=\Omega=-gA_{t}^{i}=-g\frac{1-H^{-1}}{\sinh{\beta}}.
\end{eqnarray}


The event horizon is located where $f=0$. The horizon radius can be
expressed as a power series in $\sinh{\beta}$. We will keep terms
only up to forth order in $\sinh{\beta}$ (or equivalently $y^2$)
\begin{eqnarray}
u_{H}=u_{0}=1+\frac{4}{3}y+2y^2+\mathcal{O}(y^3).
\end{eqnarray}
 The Hawking temperature associated to this horizon is given by
\begin{eqnarray}
T_{H}&=&\frac{\partial_{r}(H^{-2}f)}{4\pi}|_{r=r_{H}},
\end{eqnarray}
which leads to
\begin{eqnarray}
T_{H}&=&3\frac{2^{-2/3}}{\pi}(\mu
g^4)^{1/3}(1-\frac{2}{3}y-\frac{5}{9}y^2+\mathcal{O}(y^3)),\\\nonumber
&=&T_{0}(1-\frac{2}{3}y-\frac{5}{9}y^2+\mathcal{O}(y^3)),
\end{eqnarray}
where $T_{0}$ is the Hawking temperature at $\Omega=0$. The
dimensionless ratio $\Omega/T_{H}$ can be expanded to the second
order in $y$ as well
\begin{eqnarray}\label{OmegaoverT}
(\frac{\Omega}{T_{H}})^2&=&-\frac{2^{4/3}\pi}{3}(\mu
g)^{1/3}\sinh{\beta}(1+y+\frac{14}{9}y^2+\mathcal{O}(y^3)),\\\nonumber
&=&\frac{4\pi^2}{9}y(1+y+\frac{14}{9}y^2+\mathcal{O}(y^3)).
\end{eqnarray}

Using above and definitions (\ref{defHf}), one can write $y$ in
terms of the ratio $\Omega/T_{H}$ up to forth order as follows
\begin{eqnarray}
y&=&\frac{9}{4\pi^2}(\frac{\Omega}{T_{H}})^2-2\frac{81}{16\pi^4}(\frac{\Omega}{T_{H}})^4+\ldots
\end{eqnarray}

\section{\small Perturbing the R-Charged AdS$_{4}$ Black Holes}
\subsection{\small Review of the Reissner-Nordstr\"om Black Hole Perturbation Theory}

What follows is a lightning review of the perturbation theory of
the 4-dimensional Einstein-Maxwell system with a cosmological
constant. We closely follow \cite{Chandrasekhar}. Coordinates are
labeled as ($t,\phi,r,\theta)=(0,1,2,3)$. We follow the mostly
minus signature convention for the spacetime metric. Perturbations
are assumed to be generically non-stationary but axially
symmetric. The most general non-stationary, axially symmetric
perturbation of an arbitrary 4-dimensional spacetime can be
parameterized as follows
\begin{eqnarray}
ds_{4}^2=e^{2\nu}dt^2-e^{2\psi}(d\phi-\omega
dt-q_{r}dr-q_{\theta}d\theta)^2-e^{2\mu_{r}}dr^2-e^{2\mu_{\theta}}d\theta^2.
\end{eqnarray}
It can be shown that the linearized perturbations fall into two
distinct decoupled classes. One set, called ``polar
perturbations'', consists of $\delta
F_{02},F_{03},F_{23},\delta\nu,\delta\mu_{r},\delta\mu_{\theta}$
while the other set called ``axial perturbations'', includes
$F_{01},F_{12},F_{13},\omega,q_{r},q_{\theta}$, where $F_{ab}$
denotes the Maxwell field strength. We use $\delta$ in front of a
fluctuation, whenever the corresponding fluctuation has a non zero
background. We will not be considering polar perturbations here
since the relevant perturbations to the viscosity computations
fall into the {\em axial perturbations} class. The equations of
motion governing perturbations are most easily written in the
tetrad basis. The explicit form of the tetrad we use here is given
by
\begin{eqnarray}\label{tetrad}
e_{\mu}^{\widehat{0}}&=&(e^{\nu},0,0,0),\\\nonumber
e_{\mu}^{\widehat{1}}&=&(-\omega
e^{\psi},e^{\psi},-q_{r}e^{\psi},-q_{\theta}e^{\psi}),\\\nonumber
e_{\mu}^{\widehat{2}}&=&(0,0,e^{\mu_{r}},0),\\\nonumber
e_{\mu}^{\widehat{3}}&=&(0,0,0,e^{\mu_{\theta}}).\nonumber
\end{eqnarray}
The hatted indices are flat. Note that all the indices refer to
the tetrad basis (\ref{tetrad}) unless otherwise mentioned.

\subsection{\small The Axial Perturbation Equations of Motion}

The axial class of perturbation equations come from the following
components of the Einstein and Maxwell's equations
\begin{itemize}
\item (ab)=(12) and (13) components of the Einstein equations,
\item $\nu = \phi$ component of the Maxwell equations,
\item Bianchi identities written for ($\phi,t,r$) and ($\phi,t,\theta$) permutations.
\end{itemize}
The total number of equations sums up to five. There are two more
equations which are redundant. In what follows, ``,'' denotes
ordinary derivative with respect to the corresponding coordinate.
The explicit form of the axial equations of motion for the
background (\ref{AdS4BH}) are written as follows
\begin{eqnarray}\label{MaxwellA}
(rf^{1/2}F_{01}),_{r}+(H^2rf^{-1/2}F_{12}),_{0}&=&0,
\end{eqnarray}
\begin{eqnarray}\label{MaxwellB}
(rf^{1/2}F_{01}),_{\theta}+(H^2r^2F_{13}),_{0}&=&0,
\end{eqnarray}
\begin{eqnarray}\label{MaxwellC}
(H^2rf^{-1/2}F_{01}),_{0}+(rf^{1/2}F_{12}),_{r}+(F_{13}),_{\theta}&=&H^2r^2F_{02}Q_{02}.
\end{eqnarray}
where $Q_{0\mathcal{A}}=\omega,_{\mathcal{A}}-q_{\mathcal{A},0}$
and
$Q_{\mathcal{A}\mathcal{B}}=q_{\mathcal{A}},_{\mathcal{B}}-q_{\mathcal{B}},_{\mathcal{A}}$
and $\mathcal{A},\mathcal{B}=1,2,3$. Taking derivatives with
respect to $r,\theta$ and $t$ in equations (\ref{MaxwellA}),
(\ref{MaxwellB}) and (\ref{MaxwellC}) yields
\begin{eqnarray}\label{MaxwellA'}
(rf^{1/2}F_{01}),_{r},_{r}+(H^2rf^{-1/2}F_{12}),_{0},_{r}&=&0 ,
\end{eqnarray}
\begin{eqnarray}\label{MaxwellB'}
(rf^{1/2}F_{01}),_{\theta},_{\theta}+(H^2r^2F_{13}),_{0},_{\theta}&=&0,
\end{eqnarray}
\begin{eqnarray}\label{MaxwellC'}
(H^2rf^{-1/2}F_{01}),_{0},_{0}+(rf^{1/2}F_{12}),_{r},_{0}+(F_{13}),_{\theta},_{0}&=&H^2r^2F_{02}Q_{02},_{0}.
\end{eqnarray}
 From (\ref{MaxwellA}) we have
 \begin{eqnarray}\label{MaxwellD'}
\large[fH^{-2}(rf^{1/2}F_{01}),_{r}\large],_{r}+(rf^{1/2}
F_{12},_{0}),_{r}&=&0.
\end{eqnarray}
Utilizing (\ref{MaxwellA'}), (\ref{MaxwellB'}), (\ref{MaxwellC'})
and (\ref{MaxwellD'}) we obtain
\begin{eqnarray}\label{EqF01}
\large[H^{-2}f(rf^{1/2}F_{01}),_{r}\large],_{r}+\frac{f^{1/2}}{H^2
r}(F_{01})_{\theta},_{\theta}-rH^2f^{-1/2}(F_{01}),_{0},_{0}&=&-H^2r^2F_{02}Q_{02},2.
\end{eqnarray}
Now, let us turn to the Einstein equations. From (12) and (13)
components of the Einstein equations, we obtain
\begin{eqnarray}\label{R12}
R_{12}&=&-\frac{1}{2}e^{-2\psi-\nu-\mu_{\theta}}\large[(e^{3\psi+\nu-\mu_{r}-\mu_{\theta}}Q_{32}),_{\theta}
-(e^{3\psi-\nu-\mu_{r}+\mu_{\theta}}Q_{02}),_{0}\large]\\\nonumber
&=&-2F_{01}F_{20}.
\end{eqnarray}
Using (\ref{AdS4BH}) and (\ref{R12}) we get
\begin{eqnarray}\label{R12FEOM}
\frac{f^{-1/2}}{H^2r^3}\large[r^2fQ_{32},_{\theta}-H^4r^4Q_{02},_{0}\large]&=&-4F_{10}F_{02}.
\end{eqnarray}
The $R_{13}$ component yields
\begin{eqnarray}\label{R13FEOM}
(r^2fQ_{23}),r&=&(H^4r^2f^{-1})Q_{03},_{0}.
\end{eqnarray}
Equations (\ref{R12FEOM}) and (\ref{R13FEOM}) simplify to
\begin{eqnarray}\label{EinsteinFinal}
\frac{1}{H^4r^4}Q,_{\theta}&=&-(\omega,_{r}-q_{r},_{0}),_{0}+\frac{4}{H^2r}f^{1/2}F_{10}F_{02},\\\nonumber
\frac{f}{H^4r^2}Q,_{r}&=&(\omega ,_{\theta}-q_{\theta},_{0}),_{0},
\end{eqnarray}
where $Q=r^2f(q_{r},_{\theta}-q_{\theta},_{r})$. Since
$\partial_{t}$ and $\partial_{\theta}$ are killing directions of the
unperturbed background, a typical fluctuation will have the
following form
\begin{eqnarray}\label{ansatz}
\xi(t,r,\theta)&=&\xi(r)e^{i\sigma t+ iq\theta},
\end{eqnarray}
where $\xi$ denotes a typical fluctuation. \footnote{``$q$''
appearing in (\ref{ansatz}) is dimensionless. Therefore to compare
our results with \cite{Herzog}, one needs to send the dimensionless
$q$ to $q/2g$.}

Eliminating $\omega$ from (\ref{EinsteinFinal}) leads to
\begin{eqnarray}\label{EqQ}
\partial_{r}\large[\frac{f}{H^4r^2}Q,_{r}\large]+\frac{1}{H^4r^4}\partial_{\theta}^2Q-\frac{1}{r^2f}\partial^{2}_{t}Q&=& \frac{4}{H^2r}f^{1/2}F_{02}F_{10},_{\theta}.
\end{eqnarray}
Now let us return to equation(\ref{EqF01}). Using
(\ref{EinsteinFinal}), one has
\begin{eqnarray}
\large[H^{-2}f(rf^{1/2}F_{01}),_{r}\large],_{r}+\frac{f^{1/2}}{H^2
r}(F_{01})_{\theta},_{\theta}+(\sigma^2rH^2f^{-1/2}-4rF_{02}^2f^{1/2})F_{01}&=&
\frac{F_{02}}{H^2r^2}Q,_{\theta}.
\end{eqnarray}
Substituting the assumed form for the fluctuations (\ref{ansatz}),
one obtains the following pair of equations
\begin{eqnarray}\label{MasterEq1}
\large[H^{-2}f(rf^{1/2}F_{01}),_{r}\large],_{r}-\frac{q^2f^{1/2}}{H^2
r}(F_{01})+(\sigma^2rH^2f^{-1/2}-4rF_{02}^2f^{1/2})F_{01}&=&
i\frac{qF_{02}}{H^2r^2}Q,\\\nonumber
\partial_{r}\large[\frac{f}{H^4r^2}Q,_{r}\large]-\frac{q^2}{H^4r^4}Q+\frac{\sigma^2
}{r^2f}Q&=&-i\frac{4q}{H^2r}f^{1/2}F_{02}F_{01}.\nonumber
\end{eqnarray}
It is also useful to work out the second order differential
equation satisfied by the fluctuation $\omega$. In order to do so,
the Einstein equation corresponding to $R_{01}$ needs to be
written down
\begin{eqnarray}
R_{01}&=&-\frac{1}{2}e^{-2\psi-\mu_{r}-\mu_{\theta}}\large[(e^{3\psi-\nu-\mu_{r}+\mu_{\theta}}Q_{20}),_{r}
+(e^{3\psi-\nu+\mu_{r}-\mu_{\theta}}Q_{30}),_{\theta}\large]\\\nonumber
&=&2F_{02}F_{12}.
\end{eqnarray}
Using the AdS black hole background fields in (\ref{AdS4BH}) and
simplifying the resulting expressions, one is led to
\begin{eqnarray}\label{R01FEOM}
\frac{f^{1/2}}{H^4r^3}\Large([H^4r^4(q_{r},_{0}-\omega,_{r})],_{r}+[H^4r^2f^{-1}(q_{\theta},_{0}-\omega,_{\theta})],_{\theta}\Large)&=&
-4F_{02}F_{12}.
\end{eqnarray}
Notice that utilizing the tetrad basis definitions in
(\ref{tetrad}), the spacetime $F_{01}$ and $F_{02}$ are given by
\begin{eqnarray}
\mathcal{F}_{01}&=&rf^{1/2}F_{01},\\\nonumber
\mathcal{F}_{02}&=&F_{02},\nonumber
\end{eqnarray}
where $\mathcal{F}$ denotes the curved spacetime $F$. Using the
convenient gauge where $q_{r}=0$, (\ref{MasterEq1}) leads to
\begin{eqnarray}\label{SpacetimeF}
\frac{d}{dr}\large[\frac{f}{H^2}\frac{d}{dr}\mathcal{F}_{01}\large]-\frac{q^2}{H^2r^2}\mathcal{F}_{01}+\frac{\sigma^2H^2}{f}\mathcal{F}_{01}
&=&i\frac{\sigma H^2r^2}{u^2}F_{02}\omega,_{r}.
\end{eqnarray}
Rewriting equation (\ref{R01FEOM}) in the gauge $q_{r}=0$ gives
\begin{eqnarray}
-(H^4r^4\omega,_{r}),_{r}+H^4r^2f^{-1}(q_{\theta},_{0},_{\theta}-\omega,_{\theta},_{\theta})&=&-\frac{4F_{02}F_{12}}{f^{1/2}}H^4r^3.
\end{eqnarray}
Simplifying the above equation utilizing definitions given in
(\ref{defHf}), we will obtain
\begin{eqnarray}\label{MasterEq2}
(H^4\omega^{'})^{'}-\frac{2}{u}H^4\omega^{'}-\frac{H^4}{A(H^4-u^3)}(\sigma
qq_{\theta}+q^2\omega)&=&\frac{4H^4R_{0}^2}{u^3f^{1/2}}F_{02}F_{12},
\end{eqnarray}
where ``prime" refers to $d/du$ and $A=4g^2R_{0}^4$. Using
(\ref{EinsteinFinal}) and (\ref{MaxwellA}) combined with well the
gauge condition and (\ref{defHf}), one ends up with
\begin{eqnarray}\label{ConstraintMaxwell}
\frac{A}{H^4R_{0}^4}(H^4-u^3)iqq_{\theta},_{r}&=&-i\sigma\omega,_{r}+\frac{4}{H^2r}f^{1/2}F_{01}F_{02},\\\nonumber
(rf^{1/2}F_{01}),_{r}&=&i\sigma H^2rf^{-1/2}F_{12}.\nonumber
\end{eqnarray}

Changing the coordinate system to ``$u$'', equation
(\ref{SpacetimeF}) is written as follows
\begin{eqnarray}\label{SpacetimeFFinal}
\frac{d}{du}\large[\frac{H^4-u^3}{H^2}\frac{d}{du}\mathcal{F}_{01}\large]+\frac{1}{A}(\frac{\sigma
^2R_{0}^4H^2}{A(H^4-u^3)}-\frac{q^2}{H^2})\mathcal{F}_{01}
&=&-i\frac{\sigma\mu\sinh{\beta}R_{0}^2}{A}\omega,_{u}.
\end{eqnarray}
Denoting $h=H^4\omega^{\prime}$ and using (\ref{MasterEq2}) and
(\ref{ConstraintMaxwell}) one ends up with a second order ODE for
$h$ which has the following form in ``$u$" coordinates
\begin{eqnarray}\label{EinsteinFinalh}
h^{''}-\large(\frac{(u^3H^{-4})^{'}}{1-u^3H^{-4}}+\frac{2}{u}\large)h^{'}+\large(\frac{2}{u^2}+\frac{2}{u}\frac{(u^3H^{-4})}{1-u^3H^{-4}}\large)h
=\\\nonumber
\large[\frac{4\sigma^2H^6R_{0}^2}{A^2(H^4-u^3)^2}F_{02}+\frac{4H^4R_{0}^2}{Au^2(H^4-u^3)}(\frac{q^2u^2}{H^2R_{0}^4}-\frac{\sigma^2
H^2}{f})F_{02}\large]A_{1}+\frac{4u^2H^2f}{A(H^4-u^3)R_{0}^2}F_{02}^{'}A_{1},_{u},
\end{eqnarray}
where, $\mathcal{F}_{01}=-i\sigma A_{1}=-i\sigma A_{\phi}$ using the
fact that $\partial_{t}$ is a killing direction and remembering that
we are considering axially symmetric perturbations.

\section{\small Solving The Coupled System of ODEs}
\subsection{\small Singularity Structure, Boundary Conditions}
The system of coupled differential equations (\ref{SpacetimeFFinal})
and (\ref{EinsteinFinalh}) forms the fundamental set of equations to
be solved. As is clear, these ODEs are singular at $u=u_{0}$ where
$u_{0}$ is the horizon location. In order to isolate the singularity
at $u=u_{0}$, we substitute the following ansatz into the above ODEs
\begin{eqnarray}\label{ansatzhF}
A_{1}&=&(u_{0}-u)^{\gamma}P(u),\\\nonumber
h&=&(u_{0}-u)^{\nu}F(u).\nonumber
\end{eqnarray}
The regularity condition, in addition to the incoming boundary
condition for the fluctuations $F(u)$ and $P(u)$ at $u=u_{0}$, will
fix the values of $\gamma$ and $\nu$ (as will be computed later)
where $u_{0}$ is the horizon radius. Substituting the ansatz
(\ref{ansatzhF})into (\ref{EinsteinFinalh}) and
(\ref{SpacetimeFFinal}) gives
\begin{eqnarray}\label{MasterEqFinalForm}
F(u)^{''}+\mathcal{P}(u)F(u)^{'}+\mathcal{Q}(u)F(u)&=&
\mathcal{R}(u)P(u)+\mathcal{S}(u)P(u)^{'},\nonumber
\end{eqnarray}
\begin{eqnarray}
P(u)^{''}+\mathcal{U}(u)P(u)^{'}
+\mathcal{V}(u)P(u)=\mathcal{W}(u)F(u),\nonumber
\end{eqnarray}
where
\begin{eqnarray}\label{ODECoeffs}
\mathcal{P}(u)&=&-\large(\frac{2\nu}{u_{0}-u}
+\frac{(u^3H^{-4})^{'}}{1-u^3H^{-4}}+\frac{2}{u}\large),\\\nonumber
\mathcal{Q}(u)&=&\large[\frac{\nu(\nu-1)}{(u_{0}-u)^2}+\frac{\nu}{u_{0}-u}(\frac{(u^3H^{-4})^{'}}{1-u^3H^{-4}}+\frac{2}{u})
+\frac{2}{u^2}+\frac{2}{u}\frac{(u^3H^{-4})^{'}}{1-u^3H^{-4}},\\\nonumber
&&-\frac{q^2}{A(H^4-u^3)}+\frac{\sigma^2H^4R_{0}^4}{A^2(H^4-u^3)^2}
-\frac{4H^2R_{0}^4}{Au^2(H^4-u^3)}F_{02}^2\large],\\\nonumber
\mathcal{R}(u)&=&\large[\frac{4\sigma^2H^6R_{0}^2}{A^2(H^4-u^3)^2}F_{02}
\\\nonumber &&+\frac{4H^4R^{2}_{0}}{Au^2(H^4-u^3)}(\frac{q^2u^2}{H^2R_{0}^4}
-\frac{\sigma^2H^2}{f})F_{02} -\frac{4\gamma
u^2H^2f}{A(H^4-u^3)(u_{0}-u)R_{0}^2}F_{02}^{'}\large],\\\nonumber
\mathcal{S}(u)&=&\frac{4u^2H^2
f}{A(H^4-u^3)R_{0}^2}F_{02}^{'},\\\nonumber
\mathcal{U}(u)&=&-\large(\frac{2\gamma}{u_{0}-u}+\frac{3u^2-4H^3H^{'}}{H^4-u^3}+2\frac{H^{'}}{H}\large),\\\nonumber
\mathcal{V}(u)&=&\large[\frac{\gamma(\gamma-1)}{(u_{0}-u)^2}+\frac{\gamma}{u_{0}-u}(\frac{3u^2-4H^3H^{'}}{H^4-u^3}+2\frac{H^{'}}{H})
+\frac{\sigma^2H^4R_{0}^4}{A^2(H^4-u^3)^2}-\frac{q^2}{A(H^4-u^3)}\large],\\\nonumber
\mathcal{W}(u)&=&\frac{\mu\sinh{\beta}R_{0}^2}
{AH^2(H^4-u^3)}.\nonumber
\end{eqnarray}As it was mentioned earlier, $\nu$ and $\gamma$ can be computed by
demanding regularity for functions $F(u)$ and $P(u)$ at $u=u_{0}$.
$\nu$ and $\gamma$ are thus given by the following expressions
\begin{eqnarray}
\gamma=\nu=\pm i\frac{\sigma
R_{0}^2}{3A}(1+\frac{2}{3}y+y^2+\mathcal{O}(y^3)).
\end{eqnarray}
Note that above expression can only be trusted to the second order
in $\sinh{\beta}$. The minus sign corresponds to the the incoming
boundary condition at the horizon and according to the prescription
for calculating Minkowskian retarded Green's functions is the right
boundary condition.
\subsection{\small Solving the System in Power Series, Domain of Convrgence}

In this subsection, we find the solution to the system
(\ref{MasterEqFinalForm}) in a series expansion form around
$u=u_{0}$. The aim will be to see if the radius of convergence of
the series is large enough to include the point $u=0$, where one is
actually interested in calculating the pole structure of the
Minkowskian Green's functions. As is obvious from the ODEs
(\ref{MasterEqFinalForm}), there exist 4 singular points: $u=0$,
$u=u_{0}$, $u=\infty$, and $u=-R_{0}^2/(\mu\sinh^2{\beta})$.
Normally the radius of convergence of a series solution as viewed on
the complex plane of $u$, extends all the way to the next
neighboring singularity. For small values of $\beta$, which is what
we are considering here, the point $u=-R_{0}^2/(\mu\sinh^2{\beta})$
will be well outside the convergence circle centered around
$u=u_{0}$ and encompassing $u=0$. So there appears to be no
obstruction to continue the expansion to $u=0$. Before presenting
the solution, let us repackage our coefficients
\begin{eqnarray}\label{SimplifiedODECoeffs}
\mathcal{Q}(u)&=&\large[\frac{\gamma(\gamma-1)}{(u_{0}-u)^2}+\frac{\gamma}{u_{0}-u}(\frac{(u^3H^{-4})^{'}}{1-u^3H^{-4}}+\frac{2}{u})
+\frac{2}{u^2}+\frac{2}{u}\frac{(u^3H^{-4})^{'}}{1-u^3H^{-4}},\\\nonumber
&&-\frac{Q^2}{(H^4-u^3)}+\frac{S^2H^4}{4(H^4-u^3)^2}
-\frac{4xu^2\sinh^2{\beta}}{H^2(H^4-u^3)}\large],\\\nonumber
\mathcal{R}(u)&=&\frac{1}{R_{0}^4}\large[4Q^2x\frac{u^2\sinh{\beta}}{H^4-u^3}-8\gamma
x\frac{u\sinh{\beta}}{(u_{0}-u)H}\large],\\\nonumber
\mathcal{S}(u)&=&8 x \frac{u\sinh{\beta}}{HR_{0}^4},\\\nonumber
\mathcal{V}(u)&=&\large[\frac{\gamma(\gamma-1)}{(u_{0}-u)^2}+\frac{\gamma}{u_{0}-u}(\frac{3u^2-4H^3H^{'}}{H^4-u^3}+2\frac{H^{'}}{H})
+\frac{S^2H^4}{4(H^4-u^3)^2}-\frac{Q^2}{(H^4-u^3)}\large],\\\nonumber
\mathcal{W}(u)&=&\frac{R_{0}^4\sinh{\beta}}{H^2(H^4-u^3)},\nonumber
\end{eqnarray}
where $x=\mu/R_{0}^2, S=\sigma/(g\sqrt{x}), Q=q/\sqrt{x}$. In the
hydrodynamic limit, we will be interested only in expansions of the
functions $F$ and $P$ at most to 3rd order in $S$ and $Q$. To be
precise; we need to keep terms proportional to $S$ and $Q^2$ and
nothing. This is because of the fact that diffusion phenomenon
always involves two derivative with respect to the spatial
dimensions, while there is only one derivative with respect to time.

\subsection{\small Numeric-Symbolic Solution}

In this section, we present our series solution to the coupled
system of ODE's for the gravitational and gauge fluctuations.

Let us digress for a moment and focus on how many integration
constants one should expect in the solution. We have two second
order ODE's, which means there are four integration constants. Two
out of four are fixed by requiring the regularity condition for
$F(u)$ and $P(u)$ at $u=u_{0}$. The remaining two integration
constants get fixed by imposing boundary condition at the boundary
of AdS, i.e., at $u=0$.  Starting from the series solution ansatz
\begin{eqnarray}\label{SeriesExpansions}
F(u)&=&\sum_{i=0}^{\infty}f_{i}(u-u_{0})^i,\\\nonumber
P(u)&=&\sum_{i=0}^{\infty}p_{i}(u-u_{0})^i,
\end{eqnarray}
 our plan will be to solve for $f_{i}=f_{i}(S,Q,y)$ and
$p_{i}=p_{i}(S,Q,y)$ up to a desired order ``N'', as a function of
the two remaining integration constants (which will turn out to be
$f_{0}$ and $p_{0}$).
$y=(9/4\pi^2)(\Omega/T_{H})^2-2(81/16\pi^4)(\Omega/T_{H})^4+\mathcal{O}((\Omega/T_{H})^6)$
is the combination introduced in subsection (4.2). These series
coefficients will be further expanded to the first order in S and
second order in Q order which are the only relevant terms in the
hydrodynamic limit
\begin{eqnarray}\label{ansatz10}
f_{i}(S,Q,y)&=&\Phi_{i0}(y)+\Phi_{i1}(y)S+\Phi_{i2}(y)Q^2,
\\\nonumber
p_{i}(S,Q,y)&=&\Pi_{i0}(y)+\Pi_{i1}(y)S+\Pi_{i2}(y)Q^2.
\end{eqnarray}

We  further expand $\Phi_{ki}=\Phi_{ki}(y)$ and
$\Pi_{ki}=\Pi_{ki}(y)$ in powers of $y$

\begin{eqnarray}\label{ansatz11}
\Phi_{ki}(y)&=&\phi_{ki0}+\phi_{ki1}y+\phi_{ki2}y^2, \\\nonumber
\Pi_{ki}(y)&=&\pi_{ki0}+\pi_{ki1}y+\pi_{ki2}y^2.
\end{eqnarray}
The interpretation of the indices is clear. In order to keep our
notations simple, we have dropped the explicit dependence of $f_{i}$
and $p_{i}$ (and consequently all other expansion coefficients) on
$f_{0}$ and $p_{0}$. Now all we are required to do will be to
compute the coefficients $\phi_{kil}$ and $\pi_{kil}$.

At this stage, we need to impose boundary conditions. To do this, we
will have to use the perturbation equation (\ref{MasterEq2}).
Remembering that
\begin{eqnarray}
h=H^4\omega^{'}=(u_{0}-u)^{\gamma}F(u),
\end{eqnarray}
and taking the $u\rightarrow 0$ limit of (\ref{MasterEq2}) give rise
to
\begin{eqnarray}\label{BCF}
(H^4\omega^{'})^{'}|_{u\rightarrow
0}-\frac{2}{u}H^4\omega^{'}|_{u\rightarrow 0}-\frac{1}{A}(\sigma
qq_{\theta}^{0}+q^2\omega^{0})&=&0,
\end{eqnarray}
where superscript ``0'' refers to the boundary values of the
fluctuations $q_{\theta}$ and $\omega$ at the boundary of the
spacetime. Notice that the full solution for $F(u)$ must go to zero
at $u=0$ in order for $F(u)$ to be a regular solution of
(\ref{MasterEq2}) \footnote{in our series solution
$F(u$=$0)=\mathcal{O}(y^3)$, which is zero since we have only kept
up to two powers of $y$ at every step of our computations.} .
Thus (\ref{BCF}) can be rewritten as
\begin{eqnarray}\label{BCFFinal}
-(H^4\omega^{'})^{'}|_{u\rightarrow
0}&=&F_{10}(S,Q,y)f_{0}+F_{11}(S,Q,y)p_{0}=\frac{1}{A}(\sigma
qq_{\theta}^{0}+q^2\omega^{0}),
\end{eqnarray}
where, we have substituted the series solution for $F$. We have also
and taken into account and explicitly indicated the fact that, all
the series coefficients are expressed as a function of $S , Q, y$ as
well as $f_{0}$ and $p_{0}$. Similarly, taking the $u\rightarrow 0$
limit of (\ref{ansatzhF}) gives
\begin{eqnarray}\label{BCP}
P(u)|_{u\rightarrow
0}&=&P_{10}(S,Q,y)f_{0}+P_{11}(S,Q,y)p_{0}=A_{1}^{0},
\end{eqnarray}
where $A_{1}^{0}$ refers to the boundary value of $A_{1}$.

Equations (\ref{BCFFinal}) and (\ref{BCP}) provide us with a system
of two linear equations for two unknown $f_{0}$ and $p_{0}$ which
fixes the integration constants $f_{0}$ and $p_{0}$ in terms of the
boundary values $q^{0}_{\theta}$ and $\omega^{0}$.

 The diffusion constant denoted by $\mathcal{D}$ is the location of the $G_{tx,tx}(q,\sigma,y)$
pole which is ultimately related to the shear viscosity through the
relation
\begin{eqnarray}\label{D_eta}
\mathcal{D}&=&\frac{\eta}{\epsilon + p}.
\end{eqnarray}
The pole is given by the dispersion relation
\begin{eqnarray}\label{Dispersion}
\sigma&=&-i\mathcal{D}q^2.
\end{eqnarray}

One can easily convince oneself that the location of the desired
pole is the zero of the determinant of a 2 by 2 matrix $\Gamma$ made
out of $P_{10}$, $P_{11}$, $F_{10}$ and $F_{11}$
\begin{displaymath}
\Gamma= \left(\begin{array}{cc} P_{10} & P_{11}\\
F_{10} & F_{11} \\
\end{array}\right).
\end{displaymath}
The condition $ det(\Gamma)=0$ gives rise to the following pole
structure
\begin{eqnarray}\label{Pole}
S&=&-i\lambda(y)Q^2+\mathcal{O}(y^3).
\end{eqnarray}
In order to compute $\mathcal{D}$, we need to switch to the
dimensionful quantities. Remembering that $S=\sigma/(g\sqrt{x})$,
$Q=q/(\sqrt{x})$, one can write (\ref{Pole}) as
\begin{eqnarray}
\frac{\sigma}{(2\pi T_{0})/3}&=&-i\lambda(y)(\frac{q}{(2\pi
T_{0})/3})^{2},
\end{eqnarray}
where $T_{0}$ denotes the Hawking temperature at zero chemical
potential. \footnote{Note that in order to compare our results at
leading order with \cite{Herzog}, we need to send $q\rightarrow
q/(2g)$.} Comparing to the dispersion relation (\ref{Dispersion}),
one deduces
\begin{eqnarray}
\mathcal{D}&=&\frac{3\lambda(y)}{2\pi T_{0}}.
\end{eqnarray}
In fact, $\lambda=\lambda(y)$ is the quantity we calculate
numerically as an expansion in $y$.
\section{\small Calculating $\eta/s$}

We proceed to calculate the shear viscosity using the relation
(\ref{D_eta}). Conformal symmetry implies $\epsilon=2p$, leading to
\begin{eqnarray}
\eta&=&3p\mathcal{D}=\frac{9p\lambda(y)}{2\pi T_{0}},
\end{eqnarray}
where we have used $\mathcal{D}=3\lambda(y)/(2\pi T_{0})$ from the
previous chapter. Note that in the grand canonical ensemble one has
\begin{eqnarray}
p&=&-\frac{\partial{\Xi_{M2}}}{\partial{V}},
\end{eqnarray}
where $\Xi_{M2}=E-TS-J\Omega$ is the Gibbs free energy for the
spinning membrane. The Gibbs free energy for M2 brane with four
angular momentum turned on, can be easily calculated
\cite{Harmark&Obers}.
 Expanding to fourth order in powers of $\Omega/T_{H}$, one gets
\begin{eqnarray}\label{pressure}
p=-\frac{\Xi_{M2}}{V}=\frac{2^{7/2}\pi^2}{3^{4}}N^{3/2}T_{H}^3\large
[1+\frac{9\times4}{8\pi^2}(\frac{\Omega}{T_{H}})^2+
\frac{27}{16\pi^4}(\frac{\Omega}{T_{H}})^4+\ldots.\large].
\end{eqnarray}

Similarly, the entropy density is given by
\begin{eqnarray}
s&=&-\frac{1}{V}\frac{\partial \Xi_{M2}}{\partial
T}=\frac{2^{7/2}\pi^2}{3^{3}}N^{3/2}T_{H}^2\large
[1+\frac{9\times4}{24\pi^2}(\frac{\Omega}{T_{H}})^2-
\frac{27}{48\pi^4}(\frac{\Omega}{T_{H}})^4+\ldots.\large].
\end{eqnarray}
Assuming
\begin{eqnarray}
\lambda(y)&=&\lambda_{0}+\lambda_{1}y+\lambda_{2}y^2+\mathcal{O}(y^3),
\end{eqnarray}
the ratio $\eta/s$ becomes
\begin{eqnarray}
\frac{\eta}{s}&=&\frac{3\lambda_{0}}{2\pi}\large(1+\frac{9}{4\pi^2}(\frac{2}{3}+
\frac{\lambda_{1}}{\lambda_{0}})(\frac{\Omega}{T_{H}})^2+
\\\nonumber
&&+\frac{81}{16\pi^4}(-\frac{5}{9}
-\frac{4}{3}\frac{\lambda_{1}}{\lambda_{0}}+\frac{\lambda_{2}}{\lambda_{0}})
(\frac{\Omega}{T_{H}})^4+\ldots\large)\\\nonumber
&=&\frac{3\lambda_{0}}{2\pi}(1+\zeta_{2}(\frac{\Omega}{T_{H}})^2+\zeta_{4}(\frac{\Omega}{T_{H}})^4+\ldots).
\end{eqnarray}

As we mentioned in the previous chapter, $\lambda$ is the quantity
we compute numerically. Thus, calculating $\lambda$ will provide us
with the corrections to the ratio at zero chemical potential i.e.,
$1/(4\pi)$.

Surprisingly, the following numerical analysis presented in the
following, illustrates that the second and fourth order corrections
to the $\eta/s$ ratio asymptotes to zero as we keep more and more
terms in the Taylor expansion for F and A$_{\phi}$. This clearly
signals a saturation of the bound even in the presence of a non zero
chemical potential.

The ratio at zero chemical potential has been calculated before to
be $1/(4\pi)$ \cite{Herzog}, therefore we expect to get
$\lambda(y=0)=1/6$ at the leading order. Of course, this only serve
as a consistency check to assure us that the numerics have been done
carefully. Below, we have plotted $\lambda_{0}=\lambda(y=0)$ versus
$N$, the number of terms in the Taylor expansions for F and
A$_{\phi}$ i.e., (\ref{SeriesExpansions})
\newpage
\begin{figure}[htb!]
\epsfig{file=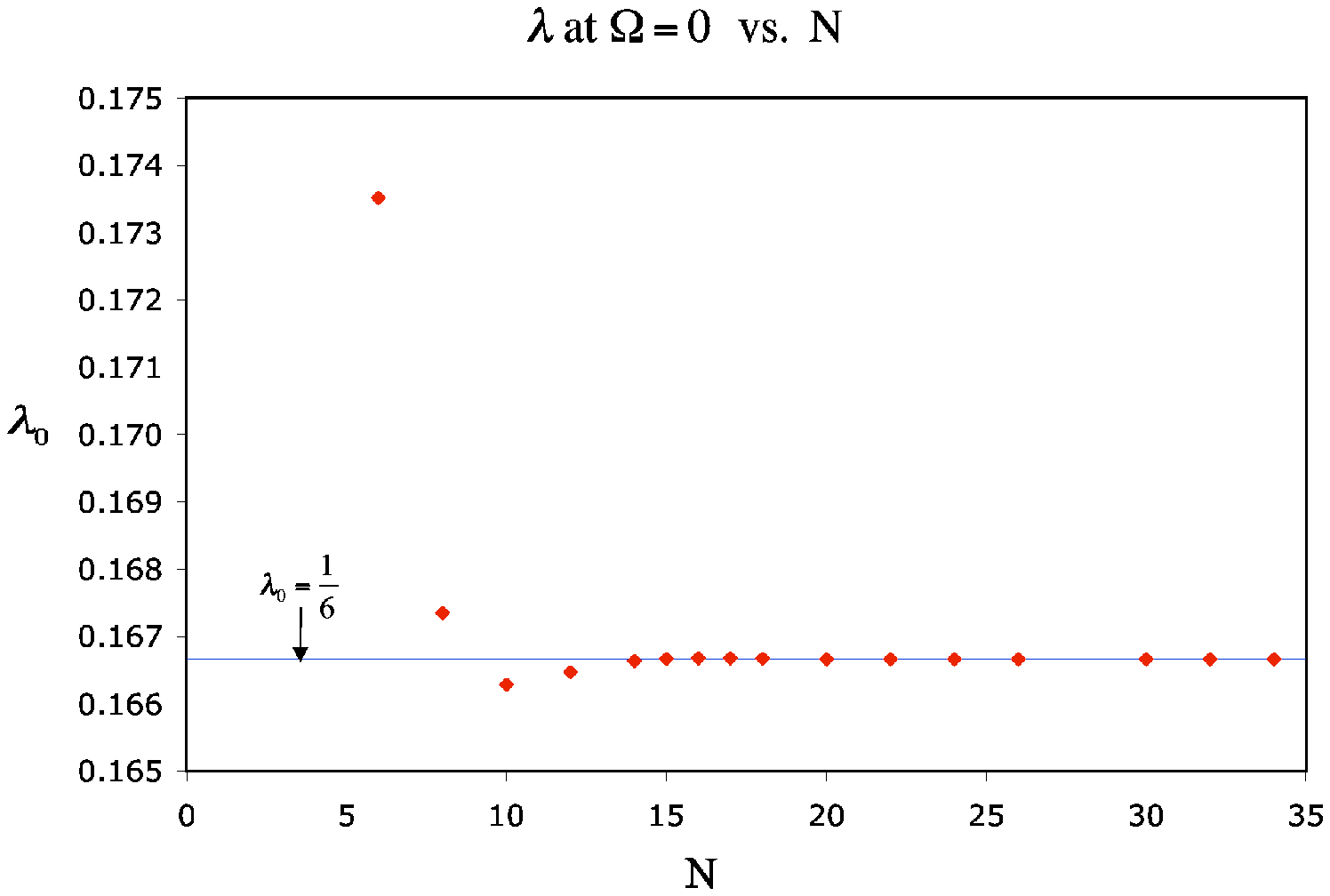, width=15cm, height=10cm}
\end{figure}
 $\zeta_{2}$, the  $(\Omega/T_{H})^2$ coefficient is the leading correction and
tends to zero as it can be easily inferred from the figure\\
\begin{figure}[htb!]
\epsfig{file=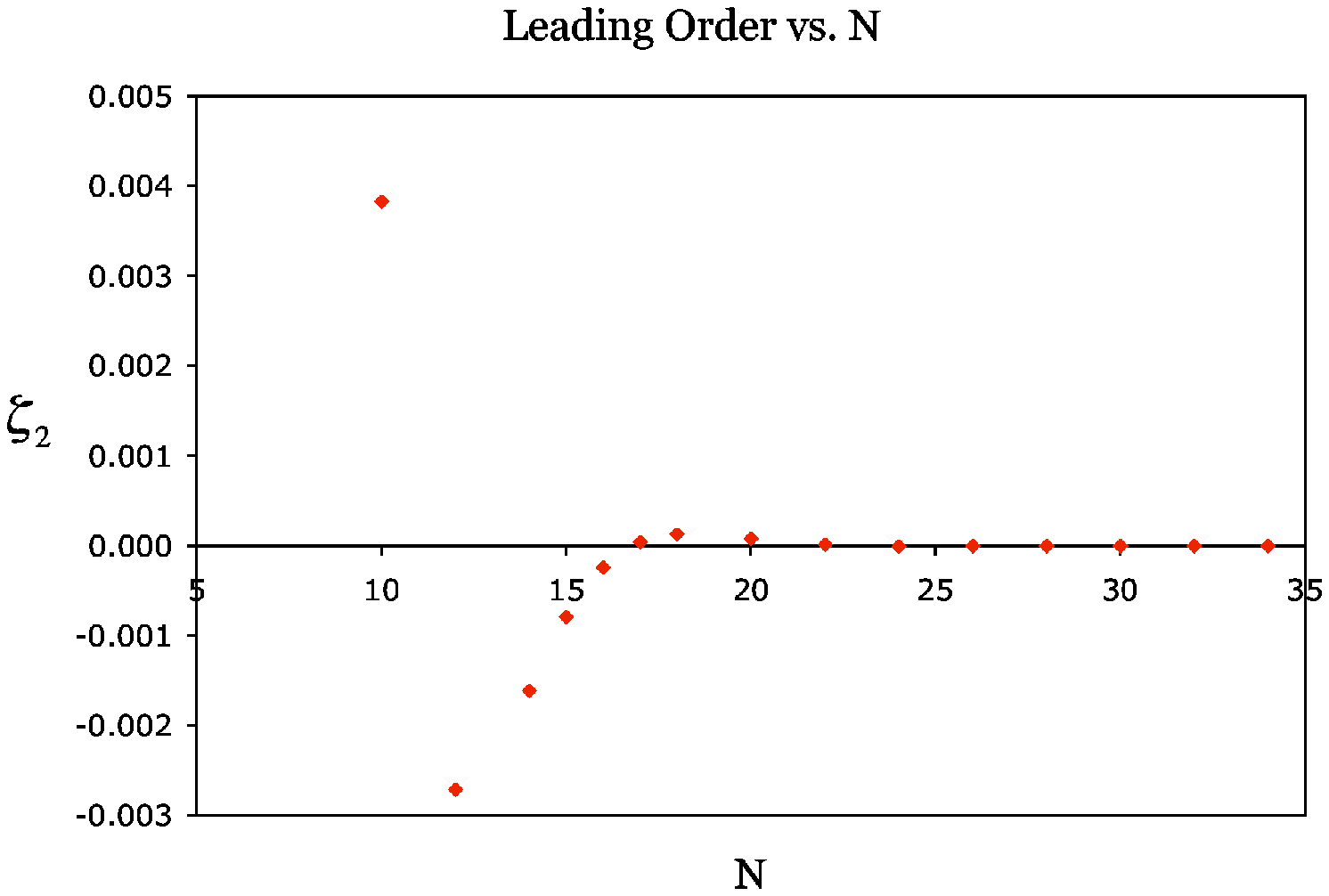, width=15cm, height=9cm}
\end{figure}
\newpage
The $(\Omega/T_{H})^4$ coefficient namely $\zeta_{4}$ also runs to
zero rather quickly as we increase N
\begin{figure}[htb!]
\epsfig{file=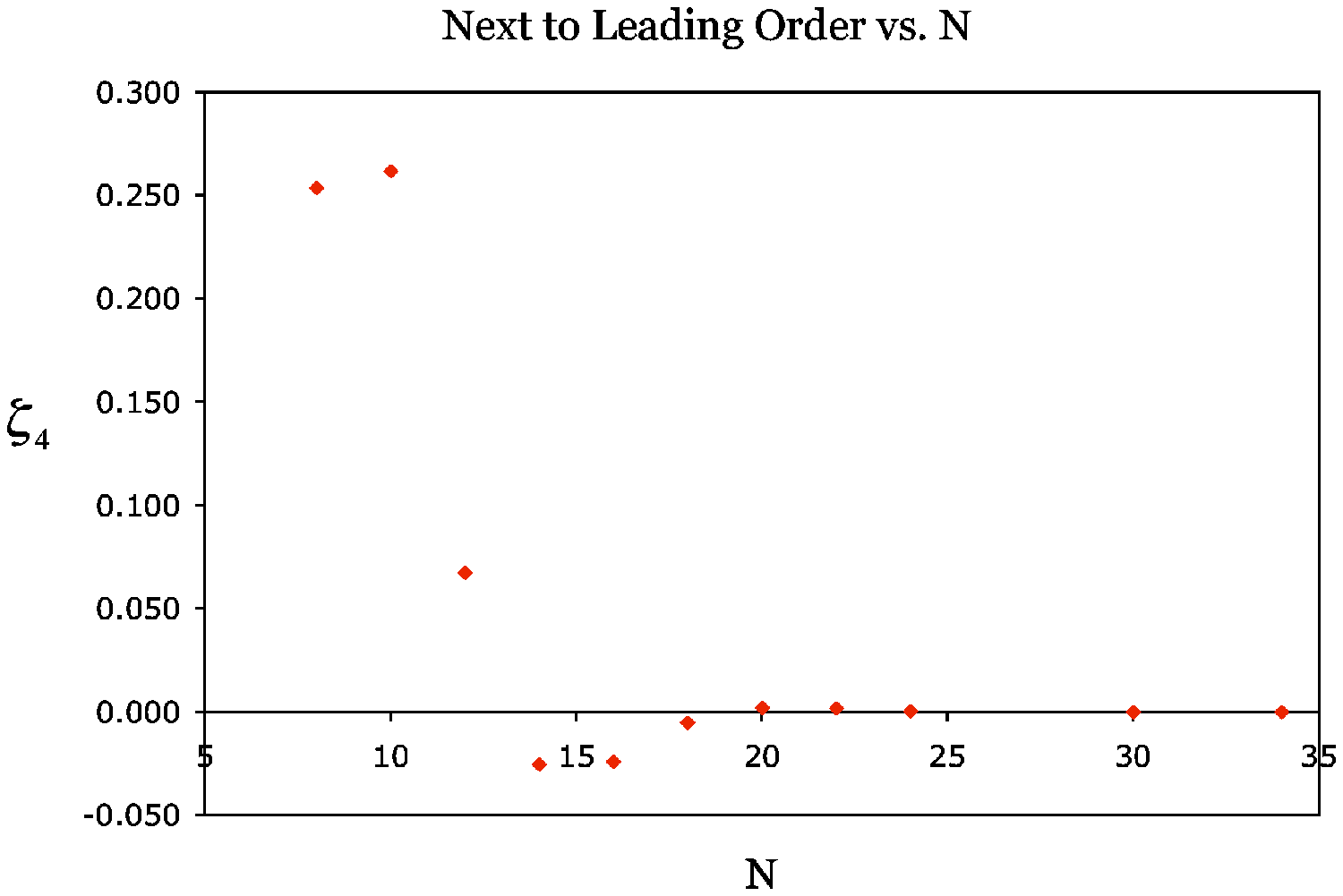, width=15cm, height=9cm}
\end{figure}
\\
\\
For completeness, we have included a shear viscosity plot versus
$\Omega/T_{H}$ for N$=30$. $\eta_{0}$ is the shear viscosity for
$y=0$
\begin{figure}[htb!]
\epsfig{file=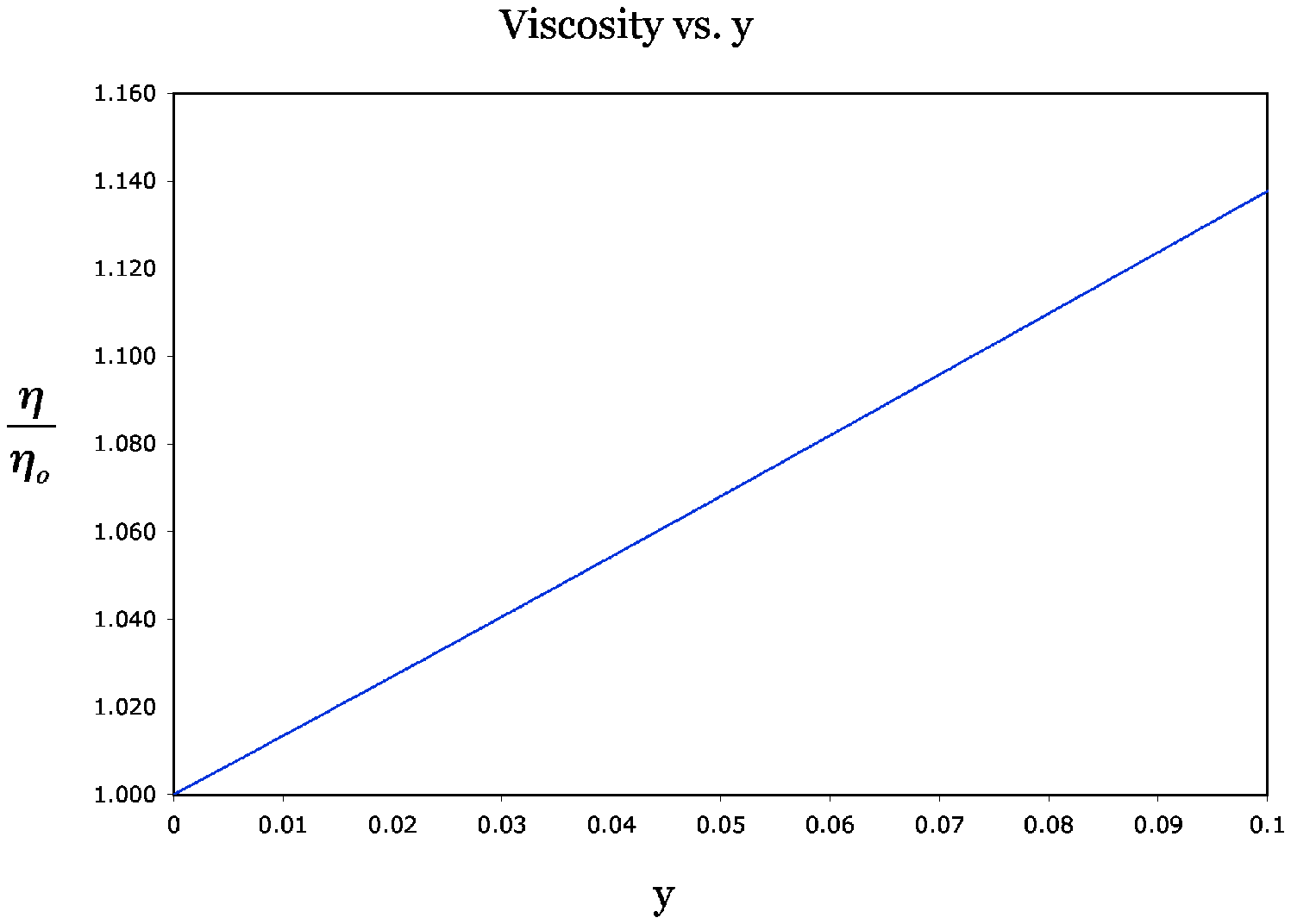, width=15cm, height=9cm}
\end{figure}

\section{\small Conclusion and Outlook}

Our results satisfies the viscosity bound conjecture proposed by
Policastro, Son and Starinets \cite{Conjecture}. The leading
finite chemical potential corrections to the viscosity itself turn
out to be positive. The ratio $\eta/s$ remained unchanged up to
fourth order in $\Omega/T_{H}$ signalling the saturation of the
bound. An interesting question is why shear viscosity increases
even though the system is literally at ``infinite coupling''.
Another way of putting is to say, what could make an ``infinitely
coupled'' system , ``less infinitely coupled'' (in order for the
viscosity to increase)! One could speculate that the reason behind
this enhancement in shear viscosity might lie in some screening
effect for the color interaction mediator at finite chemical
potential. At non zero chemical potential i.e., when the number of
various particle species carrying different R-charge is
imbalanced, the ``gluons" mass receives correction from the
chemical potential which could result in screening . This
screening of the color charge weakens the effective interactions
in the plasma which ultimately leads to a bigger mean free path.
When this paper was being written, I became aware of two other
works \cite{Private_Comm1}, \cite{Private_Comm2} in preparation on
AdS$_{5}$ with similar results. While for the membranes (which
were studied in this paper), no gauge theory description exists,
for the AdS$_{5}$ system there is a gauge theory living on the
world-volume of D3-branes. Using conformal invariance of N$=4$ SYM
theory, only based on dimensional grounds, one could argue that
the gluons mass receives corrections which are proportional to
$\mu$ where $\mu$ is the chemical potential \footnote{$\mu$ has
dimension mass. Note that, this effect is on top of the usual
finite temperature corrections}. A similar situation occurs in
perturbative QCD. In this case, one could speculate with more
confidence that the screening effect may be the true reason for an
increase in the shear viscosity of a hot gauge theory plasma at
finite chemical potential.

It was argued and quantified by Karch \cite{Karch} that the
conjectured viscosity bound is connected to Bousso's Generalized
Covariant Entropy Bound (GCEB). Given such an interesting
interrelation, one could reinterpret the viscosity bound as a {\em
non-gravitational} and {\em empirical} window to the realm of {\em
quantum gravity}. It turns out that the viscosity bound is exactly
what matter is required to obey in order for gravity to modify the
light-sheets (motion of the viscous fluid results in a stress
$T_{ij}$-generated curvature) to prevent the GCEB from a
catastrophic violation. The current formulation of the GCEB suffers
from a number of problems including ``the species problem''. The
species problem is the simple statement that the entropy of a system
of field(s) confined in a region of space can grow simply by
increasing the number of particle species while keeping the total
energy fixed. This would lead to a violation of the GCEB. An
exciting question \cite{Carlos} would be to ask whether violating
the GCEB through the species problem, for instance, would lead to a
violation of the viscosity bound. To address this question one has
to perform similar calculations as outlined in the present paper for
supergravity duals to the gauge theories with large $N_{f}$
\cite{KarchKatz} (see also \cite{Kirsch}) at ``finite temperature''.
A zero temperature realization was considered in \cite{Kirsch},
where the field theory corresponding to the localized D2-D6
intersection is an $\mathcal{N}=4$, $d=3$ super Yang Mills gauge
theory coupled to $N_{6}$ hypermultiplets in the fundamental of the
gauge group, where $N_{6}$ is the number of D6-branes. $N_{6}/N_{2}$
is kept fixed while $N_{2}, N_{6} \gg 1$. The full supergravity
background (i.e., D6-flavor branes including back-reaction) has been
worked out in \cite{Cherkis}, where the fact that uplifted D6-branes
to M-theory has a Taub-NUT space component proves to be helpful. The
non-extremal version of \cite{Cherkis} is not yet known and it seems
like a daunting task to carry out. It would be interesting to find
the background at least in the form of an expansion series in the
vicinity of the horizon. Extensions of similar sets of computations
would teach us a lot about whether there is a violation of the
viscosity bound at large $N_{f}$.

Needless to say, finding analytic solutions to the coupled
differential equations here would be invaluable as it could reveal
analytic structure of the viscosity as a function of $\Omega/T_{H}$.

There is no G$_{N}$ in the statement of the bound. This simply
indicates that the conjecture is an statement about quantum
mechanical matter without any reference to gravity. So it is natural
to expect the existence of a proof or counterexample for the bound
which only involves weakly gravitating quantum physics. The
relevance of gravity seems to be solely a consequence of the fact
that in order to get down to the saturation limit of the bound, one
is required to go to exceedingly high values of the coupling which,
in the light of AdS/CFT, is mapped to the physics of highly
gravitating objects, i.e., black holes in Anti de Sitter space.

\section{\small Acknowledgement}

I am grateful to Carlos Nunez for interesting discussions. I am
indebted to Leonard Susskind and John McGreevy at Stanford Institute
for Theoretical Physics (SITP), where the final stage of this work
was completed, for inspiring discussions and their hospitality. I
would also like to thank the PiTP summer school July 2005 organizers
at the Institute for Advanced Study in Princeton, where this work
was initiated and partially done, for providing stimulating
environment. Special thanks go to Amanda Peet and Erich Poppitz at
the University of Toronto and to Andrei Starinets and Alex Buchel at
Perimeter Institute, for discussions. OS work was supported in part
by an Ontario Graduate Scholarship (OGS) and by NSERC of Canada.

\end{document}